\documentstyle[prbbib,aps]{revtex}
\input epsf
\begin{document}
{\Large
\draft
\title{
LOW--TEMPERATURE PHASE TRANSITION IN 3--STATE POTTS GLASS
}
\author{
N. V. Gribova, V.N.Ryzhov, and E.E.Tareyeva
}
\address{
Institute for High Pressure Physics, Russian Academy of Sciences, Troitsk
142190, Moscow region, Russia}
\date{\today}
\maketitle
\begin{abstract}
The low-temperature instability of one-step replica symmetry
breaking (1RSB) phase in 3-state Potts spin glass is obtained
explicitly. The temperature of the instability is higher than
the temperature where the 1RSB entropy becomes negative.
The conjecture of the possibility of the low-temperature
full RSB is supported.
\end{abstract}

During last decades the so called 1RSB (one--step
replica--symmetry breaking) models: the $p$--state Potts spin
glasses and $p$--spin spin glasses as well as their soft--spin
versions -- are in the focus of investigations in spin glass
theory.  It is usually believed that the 1-RSB solution if
stable in the vicinity of its appearance remains stable till
zero temperature. In this paper we demonstrate
explicitly the low--temperature
instability of 1RSB solution for the 3-state Potts spin glass
using its representation in terms of quadrupole operators.

The $p$-state Potts spin glass model is a lattice model where
each lattice site carries a Potts spin $\sigma_i$ which can take
one of the $p$ values $\sigma_i = 0, 1, ..., p-1$ and the
interaction Hamiltonian is
\begin{equation}
H=- \frac{p}{2} \sum_{i \not= j}  J_{ij} \delta_{\sigma_i
\sigma_j}
\label{ham}
\end{equation}
where $\delta_{\alpha \beta}$ is the Kronecker symbol. Thus, a
pair $\{\sigma_i,\sigma _j\}$ contributes an energy $-J_{ij}$
if $\sigma _i = \sigma _j$ and zero otherwise. The interactions
$J_{ij}$ are quenched random variables described by a Gaussian
distribution
$$P(J_{ij}) = ( \sqrt {2 \pi} J)^{-1} \exp
{[-(J_{ij}-J_0)^{2}/2J^2]}.$$

The Potts glass with an infinite--range interaction
$J_0= \tilde J_0/N$, $J= \tilde J/N^{1/2}$ has been studied in
~\cite{EldSh83,LaEr83,Eld84,LaNu84,gks,LT1,LT2,LT3,LT4,sapari94}.
The short--range version has been considered in
~\cite{Gold85,cwi1,cwi2} and is a subject of intense
investigation through computer simulations ~\cite{Binder}.
The soft--spin version  of Potts glass has also been suggested
as a starting point for a theory of structural glasses and the
transition from the metastable fluid to the glass state
~\cite{kirkp}. The Potts glass may also serve as a model for
orientational glasses in molecular crystals and cluster glasses
where a strong single--site anisotropy restricts the orientation
of the appropriate molecular group to $p$ distinct directions.

The 3-state Potts spin glass is somehow intermediate system
between Sherrington-Kirkpatrick (SK) glass ($p$=2) and "canonical" 1RSB glasses ($p\ge4$).
In 3-state PG there is no reflection symmetry and 1RSB solution
was shown ~\cite{gks,LT3} to be stable in the vicinity of the
RS transition temperature (which coincides with that of
1RSB transition) against the higher stages of RSB, but the static
transition is continuous, which is not general properties of 1RSB models.

It is usually believed that the 1RSB solution if stable in the
vicinity of its appearance remains stable till zero
temperature. This fact was established rigorously by
Crisanti and Sommers ~\cite{som} in the case of the spherical
$p$--spin interaction spin glass model. The spin--glass phase of
the spherical $p$-spin spin model is
described exactly by a step order parameter function, i.e.
the 1RSB is the most general solution within the Parisi RSB
scheme.

As far as we know there is only one paper where the phase
transition from 1RSB to full RSB (FRSB) phase was
established. Using perturbations around
known solutions for the cases of $p=2+\epsilon $ and $p\to\infty$
glass E.Gardner ~\cite{gard} showed for Ising $p$-spin glass that
1RSB solution is unstable at {\it very} low temperature.
The second transition leads to a phase described by a continuous
order parameter of FRSB function $q(x)$.

The RSB solution for the mean field 3-state PG was considered in
~\cite{gks,LT2,sapari94}. It was shown ~\cite{gks,LT2} that 1RSB
solution is stable in the vicinity of phase transition. In
~\cite{gks} it was supposed that at lower temperature another
spin glass phase appears which differs from the 1RSB in the
nature of the correlations among the many degenerate ground
states of the system. However, this second phase transition was
not found in the thorough investigation by De Santis, Parisi and
Ritort ~\cite{sapari94}. There are no observations whatsoever
that would indicate that the short--range system also has two
successive phase transitions ~\cite{cwi1,cwi2,Binder}.
It is worth to notice that this possible second transition (often called
Gardner transition) to low-temperature FRSB phase was
usually regarded as an inessencial, and somehow exotic
phenomenon. However, in ~\cite{lote} it was shown that the
metastable states which are relevant for the
out--of--equilibrium dynamics of such systems are always in
a FRSB phase. This renewes the interest to the low--temperature
behavior of the systems without reflection symmetry.

Let us consider now the system of particles on lattice sites
$i,j$ with the Hamiltonian ~\cite{LT1,LT2}:
\begin{equation}
H=- \frac{1}{2} \sum_{i \not= j}  J_{ij} (Q_i Q_j + V_i V_j),
\label{ham1}
\end{equation}
where $Q=3 J^2_{z} - 2$, ${\bf J}=1, J_z=1, 0,-1; Q^2=2-Q,
V^2=2+Q, QV=VQ=V$.
A particle quadrupole moment is the second-rank tensorial
 operator with five components. In the principal axes frame only
two of them remain: $Q$ and $V$. In the subspace $J=1$ the
following equality holds:
$$ \frac {1}{6} (Q_m Q_n + V_m V_n + 2) = \delta _{mn}.$$
This equality shows the equivalence of ~(\ref{ham1}) to the
$p=3$ Potts Hamiltonian ~(\ref{ham}).
We shall assume that
$J_{ij}$ are
distributed following the
Gaussian law with zero mean:  $$P(J_{ij}) = ( \sqrt {2 \pi}
J)^{-1} \exp {[-J_{ij}^{2}/2J^2]},$$ and $J= \tilde J/N^{1/2}$.

Using the standard procedure of the replica method,
we get the expression for the free energy, corresponding
to the Hamiltonian (\ref{ham1}) \cite{LT1}:
$$\frac{\langle F\rangle_J}{NkT} = - \lim_{n \rightarrow 0} \frac{1}{n}
max \left\{ -2t^2 - t^2 \sum_{(\alpha
\beta)} ( q^{\alpha \beta} )^2 - \frac{t^2}{2} \sum_\alpha (
x^\alpha)^2 +\right.$$
\begin{equation} +\ln Tr \exp \left[
 t^2 \sum_{(\alpha \beta)} q^{\alpha \beta}
 (Q^\alpha Q^\beta + V^\alpha V^\beta) + t^2
 \sum_\alpha Q^\alpha x^\alpha \right] \label{F1}
 \end{equation}
 Here $( \alpha \beta )$ means the sum over the couples of replicas, $n$
-- the number of replicas, and $x^\alpha = \langle \langle
Q^\alpha \rangle \rangle$, $q^{\alpha \beta} = \frac{1}{2}\langle
\langle Q^\alpha Q^\beta + V^\alpha V^\beta \rangle
 \rangle$ are the order parameters.

The RS solution of saddle--point equations gives $x=0$, $q\neq0$
 for $T<T_c$ where $kT_c=2J$ ~\cite{LaEr83,LT1}. This solution is
unstable against 1RSB at $T_c$. Following Parisi scheme we carry
out the first step of RSB by dividing the $n$ replicas into
$n/m$ groups of $m$ replica and setting $q_{\alpha \beta }=q_1$
if $\alpha $ and $\beta $ belong to the same group and
$q_{\alpha \beta }=q_0$ otherwise. In the limit $n\to 0$
the parameter $m$ is constrained to the range $0\le m \le1$.
In the absence of an external field only two variables remain
$q-q_0=v$ and $m$ and the free energy takes the form:
\begin{equation}
\frac{\langle F\rangle_J}{NkT} = - 2t^2
+ \frac{t^2}{2} (m-1)v^2 + 2 t^2 v - \frac{1}{m}
\ln \int_{-\infty}^{\infty} \int_{-\infty}^{\infty} dz^G dy^G
\Psi^m \left( \theta_1,
\theta_2) \right), \label{F2} \end{equation} where $$ da^G =
\frac{1}{\sqrt {2\pi}} da e^{-a^2/2}, $$ $$ \Psi =  e^{-2
\theta_1} + e^{\theta_1}(e^{\theta_2} + e^{-\theta_2}),$$
$$\theta_1 = tz\sqrt{v},~~~\theta_2 = ty\sqrt{3v},$$
$v,m$ satisfy the saddle--point equations
(in fact, the maximum conditions) for ~(\ref{F2}):

\begin{equation}
v = \frac{1}{2} \frac{<\Psi^{m-2} \left(
(\Psi'_1)^2 + 3 (\Psi'_2)^2\right)>} {<\Psi^m>},
\label{pv}
\end{equation}

\begin{equation}
-\frac{t^2 v^2 m^2}{2}  =
 \ln < \Psi^m> -m \frac{< \Psi^m \ln \Psi> }{ < \Psi^m>}.  \label{m}
\end{equation}
Here
$$ \Psi'_i = \frac {\partial \Psi}{\partial\theta _i}$$ and
$$ <...> = \int_{-\infty} ^{\infty} \int_{-\infty}^{\infty} dz^G dy^G...$$
It is worth to notice that our free energy ~(\ref{F2})
coincides with that given by Eq.(16) of Ref.~\cite{sapari94} if
we put there $p=3$, $q=v/2$, $\beta =2 t$, introduce new variables
$y_1=z_1, y_2 = z_2 - z_1, y_3=z_3-z_1$. This enables one to intergrate
explicitly over $y_1$ so that only two integrals remain. This
simplier form increases the precision of calculations
significantly.

\begin{figure}[htb]
\begin{center}
\leavevmode
\epsfxsize = 10.0truecm
\epsfysize = 8.0truecm
\epsffile{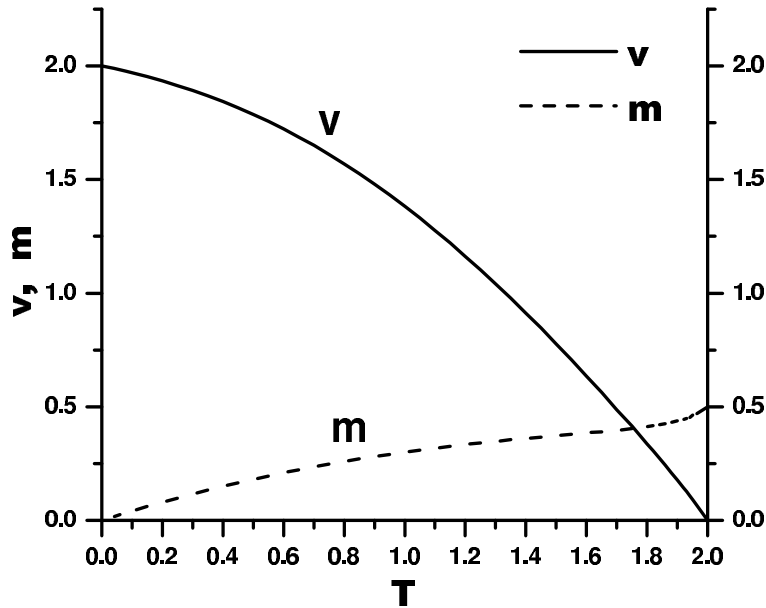}
\caption{ Order parameters as functions of {\bf\sf T}$=\frac{kT}{J}$.
(One must notice that $kT_c=2J$)}
\end{center}
\end{figure}
The solution of Eqs.~(\ref{pv})-~(\ref{m}) close to $T_c$
in the form of step--function $v(m)$ was obtained in
~\cite{gks,LT2} and was shown to be stable near $T_c$ against
further RSB. Now performing numerical maximization of ~(\ref{F2})
we obtain the values of $v$ and $m$ for all temperatures. The
results are presented in Fig.1 and do coincide with that of the
paper ~\cite{sapari94} (as well as the results for dynamical
transition temperature defined from the marginal stability
condition). The corresponding free energy ~(\ref{F2})
changes the sign of the slope at low temperature, so that the entropy
of the 1RSB solution becomes negative (see Fig.2).
To obtain the point where further RSB is necessary
one has to consider the stability of the obtained 1RSB solution
for $v$ and $m$ with respect to further replica breaking.
Following Almeida and Thouless
~\cite{AT}, we expand the free energy ~(\ref{F1}) around the 1RSB free
energy ~(\ref{F2}). Stability requires that all eigenvalues of
the stability matrix associated with fluctuations evaluated
within 1RSB solution should be positive. In the limit $n\to0$
we get six eigenvalues ~\cite{LT2}:
\begin{equation}
\lambda _1 = P + 2 (m-2) Q + 1/2 (m-2) (m-3) R;
\label{l1}
\end{equation}
\begin{equation}
\lambda _2 = P' + 2 (m-1) Q' + (m-1)^2 R';
\label{L2}
\end{equation}
\begin{equation}
\lambda _3 = P + (m-4) Q - (m-3) R;
\label{L3}
\end{equation}
\begin{equation}
\lambda _4 = P' - 2 Q' + R';
\label{L4}
\end{equation}
\begin{equation}
\lambda _5 = P - 2 Q + R;
\label{L5}
\end{equation}
\begin{equation}
\lambda _6 = P' + (m-2) Q' + (1-m) R';
\label{L6}
\end{equation}
where
$$P = 4 - 2 t^2[8 + 4v - 4v^2];$$
$$Q= -2 t^2[4v -4v^2 + t_3];$$
$$R= -2 t^2[ -4v^2 + r_4];$$
$$P'=4-16t^2;\,\, Q'=-8t^2v;\,\, R'=-4t^2 v;$$
$$t_3= - \frac {<\Psi^{(m-3)} (\Psi'_1)^3>}
{<\Psi^m>} + 9 \frac {<\Psi^{(m-3)} \Psi'_1 (\Psi'_2)^2>}{<\Psi^m>};$$
$$r_4=\frac {<\Psi^{(m-4)} [(\Psi'_1)^2 +3 (\Psi'_2)^2]^2>}{<\Psi^m>}.$$
Five of these eigenvalues occur to be always positive.
But $\lambda_5$ -- the replicon mode -- changes sign at the
temperature higher than that of the changing of the slope of
the 1RSB free energy:
\begin{equation}
\lambda _{repl} (= \lambda _5) = 4 - 2t^2[8-4v+r_4-2t_3].
\label{LR}
\end{equation}

\begin{figure}[htb]
\begin{center}
\leavevmode
\epsfxsize = 10.0truecm
\epsfysize = 8.0truecm
\epsffile{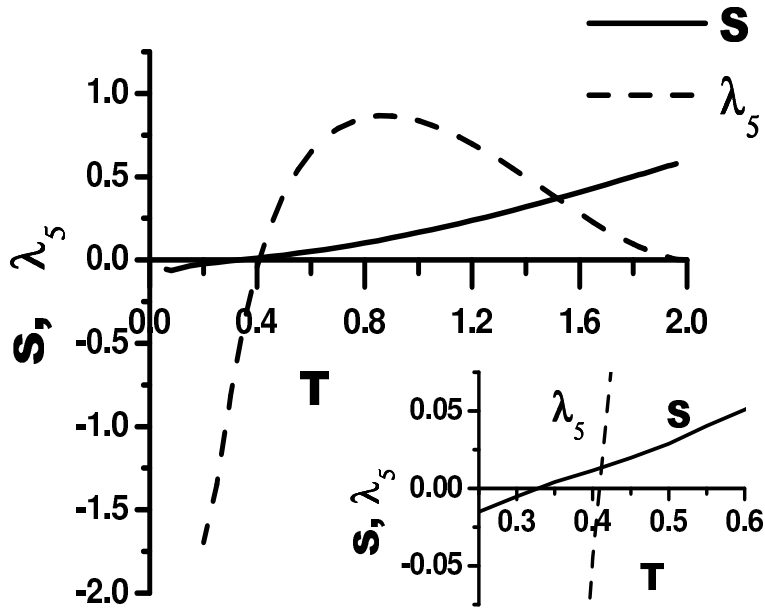}
\caption{ $\lambda _{repl}$ and the entropy as functions of
temperature ({\bf\sf T}$=\frac{kT}{J}$)}
\end{center}
\end{figure}

In Fig.2 the behavior of the $\lambda _{repl}$ along with
the entropy as functions of temperature are presented.
The instability point is $t_2 = 0.4$ and in this point the
critical values of parameters are $v_c= 1.84$ and $m_c=0.15$.

At the point $t_2$ it is necessary to perform further replica
symmetry breaking and a transition to a different type of glass
phase occurs. Further steps of RSB within Parisi scheme will
result in a new transition between the 1RSB regime and a more
complex regime. These two phases differ in the nature of the
correlations among the many degenerate ground states of the
system. We suppose that this complex regime at lower temperature
can in fact be properly described by the FRSB Ansatz by Parisi
as it was proposed in ~\cite{gks}.However, we do not exclude the
possibility of obtaining the FRSB regime not immediately at the
1RSB instability point but through a consequence of higher order
separate RSB phase transitions.

As some kind of indirect indication that zero temperature phase of
our model (\ref{ham1}) corresponds to FRSB the results of the
paper ~\cite{LT3} can be considered. The well-known result by
Edwards and Tanaka ~\cite{taned} for the number of metastable
states at zero temperature in SK spin glass is generalized there.
In Ref.~\cite{taned} the macroscopically large number $<N_s>_{SK}$
of the metastable states at $T=0$ was obtained:
$$<N_s>_{SK}=\exp [ - N \Omega _{SK}] $$
where $\Omega _{SK}=-0.19923$. This result was obtained without
any reference to RSB scheme and even replica approach. Using the
method of Ref.~\cite{taned} in the paper~\cite{LT3} the number of
metastable states $<N_s>_{3P}$ at $T=0$ for the model
(\ref{ham1}) was obtained
$$<N_s>_{3P}=\exp [ - N \Omega _{3P}] $$
with $\Omega _{3P}=\Omega _{SK}+\ln (\frac{3}{2})$,
so that the "relative" number of metastable
states (the part of all possible $p^N$ states) is the same as in
SK model :
\begin{equation}
\frac{\exp [ - N \Omega _{3P}]}{3^N} = \frac{\exp [ - N \Omega _{SK}]}{2^N}.
\label{met}
\end{equation}
It seems us that this fact shows a similarity of the structure of
zero temperature landscapes in these two models and supports the
importance to look for FRSB phase in 3-state PG.

To conclude, in this paper, using the quadrupole representation ~(\ref{ham1})
for the 3-state PG, we obtained explicitly the point of the
instability of 1RSB solution. So,
 we give some additional support to
the Gross, Kanter, Sompolinsky conjecture about the low
temperature behavior of the 3-state Potts spin glass.
 We think that our success is based on the fact that we have
fewer ploblems with precision at low temperatures, because
the equations ~(\ref{pv}) -~(\ref{m}) are simplier than the
corresponding equations of Ref.~\cite{sapari94}.

Authors thank T.I. Shchelkacheva and N.M.Chtchelkachev for
helpful discussions and valuable comments.

This work was supported in part by Russian Foundation for
Basic Researches (Grants No. 02-02-16621 (NVG and EET)
and No. 02-02-16622 (VNR)).

\end{document}